\documentclass{moriond}
\usepackage{amssymb}
\usepackage{caption}

\def\be{\begin{equation}}
\def\ee{\end{equation}}
\def\bea{\begin{eqnarray}}
\def\eea{\end{eqnarray}}

\newcommand{\Photo}{\includegraphics[height=35mm]{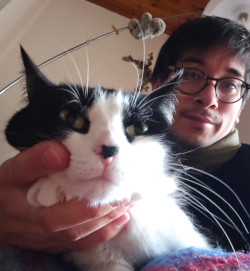}}

\begin{document}
\vspace*{4cm}
\title{Primordial high energy neutrinos}

\author{Nicolas Grimbaum Yamamoto}

\address{Service de Physique Th\'eorique, Universit\'e Libre de Bruxelles,
Boulevard du Triomphe, CP225, 1050 Brussels, Belgium}

\maketitle\abstracts{Among the few ways to probe the early Universe, neutrinos offer a particular window on high energy phenomena occurring before
recombination. We discuss the opportunities of observing \emph{primordial high
energy neutrinos} (Phenus): neutrinos produced before or around recombination from the
decay or annihilation of long-lived relics, arriving at detectors today with energies
in the GeV--PeV range. We summarise the results of a general study of this
scenario, covering the sharp spectral features such fluxes would
display, the theoretical (BBN and CMB) and experimental constraints on the source
particle parameter space, and the regions that could realistically be probed by current
and future neutrino telescopes. We also present a dedicated Monte Carlo code for
computing the distortion of the Phenu spectrum by final state radiation and interactions
with the cosmic neutrino background during propagation, and apply it to assess the
primordial origin hypothesis for the KM3-230213A ultrahigh energy neutrino
event.}

\section{Primordial high energy neutrinos}

Setting gravitational waves aside, there are only a handful of ways to probe the early
Universe from the observation of fluxes of Standard Model particles. CMB photons probe recombination at
$z \sim 1100$; Planck data have made this a cornerstone of precision cosmology, and it
is the earliest signal accessible through photons. At latter times, the 21\,cm signal probes reionisation at redshift
$z \sim 20$ and is being actively observed.  At earlier times, the cosmic
neutrino background (C$\nu$B) probes neutrino decoupling around $z \sim 10^{10}$, but because these
neutrinos carry very little energy today and interact only weakly, direct detection is
extraordinarily difficult.

This raises a natural question: could we observe neutrinos of primordial origin if they
were energetic enough to be detected by existing telescopes? This is possible if they
were produced from the decay or annihilation of a long-lived relic particle \emph{P}, at temperatures below
the neutrino decoupling temperature ($T_{\rm dec}^\nu \simeq 1$\,MeV), so that they do
not thermalise with the plasma. We call these \emph{primordial high energy neutrinos},
or Phenus.

Surprisingly, this
possibility has been relatively little studied. Ref.~\cite{Frampton:1980} put forward the
idea of a source particle decaying into 2 neutrinos and Ref.~\cite{Gondolo:1993} studied
it further from the perspective of experimental constraints.
Refs.~\cite{Kanzaki:2007,Ema:2014a,Ema:2014b,McKeen:2018,Berghaus:2019,Jaeckel:2021}
considered the scenario in specific contexts and energy ranges, notably in connection
with PeV neutrinos observed by IceCube~\cite{IceCube:2013}.

\section{Sharp spectral features and parameter space}

\subsection{Types of sharp spectral features}

\begin{figure}
    \centering
    \includegraphics[width=0.4\linewidth]{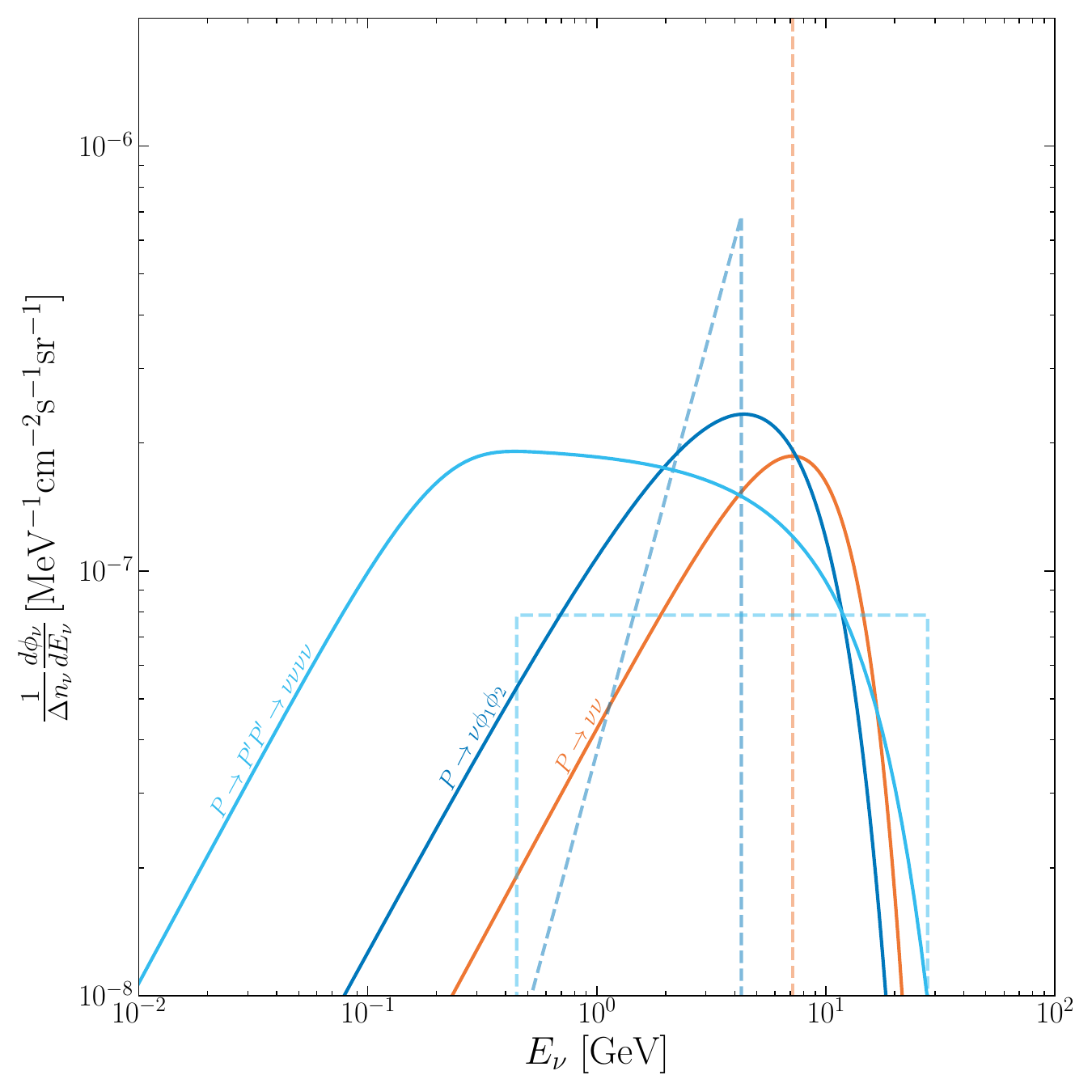}
    \includegraphics[width=0.4\linewidth]{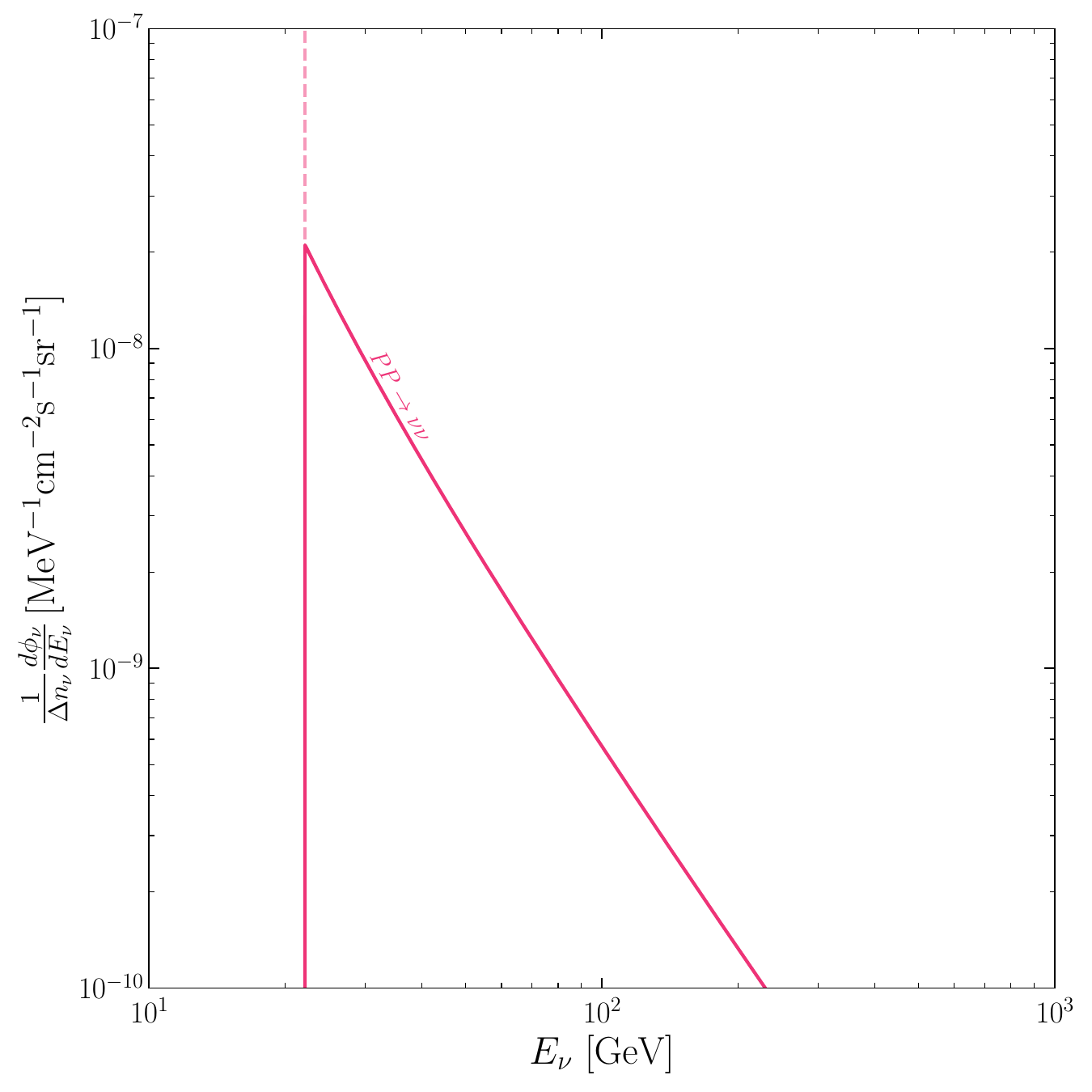}
    \caption{Neutrino flux as a function of the observed energy today for the 4 types of sharp spectral
features considered. The left panel gives it for two-body (orange), three-body (blue) and box-shaped (cyan) decays. The right panels give it for an example of an out-of-equilibrium annihilation. Also given (in dashed lines) are the original spectra
at source, rescaled in energy by an arbitrary factor (which we take so that the energy spectrum maxima
coincide, explicitly showing the effect of redshift on the spectrum).}
    \label{fig:sharp}
\end{figure}
We performed a general systematic study of this scenario in Ref.~\cite{GY:2025}.
A key observation is that even though the source decays over an extended period of
cosmic time, so that neutrinos produced at different times are redshifted by different amounts, the
resulting spectrum remains \emph{sharp} today. Four types of sharp spectral features arise depending on the production mechanism: a
two-body decay $P \to \nu\bar\nu$, a three-body decay $P \to \nu P' P''$, a box-shaped
spectrum from a cascade $P \to P'P' \to \nu\nu\bar\nu\bar\nu$, and out-of-equilibrium
annihilation $p\bar p \to \nu\bar\nu$ triggered e.g.\ by a phase transition.
The resulting energy spectra at the detector taking the redshift effect into account, for each of these four cases, are shown
in figure~\ref{fig:sharp}.
\subsection{Constraints and observable regions}

\begin{figure}
    \centering
    \includegraphics[width=0.4475\linewidth]{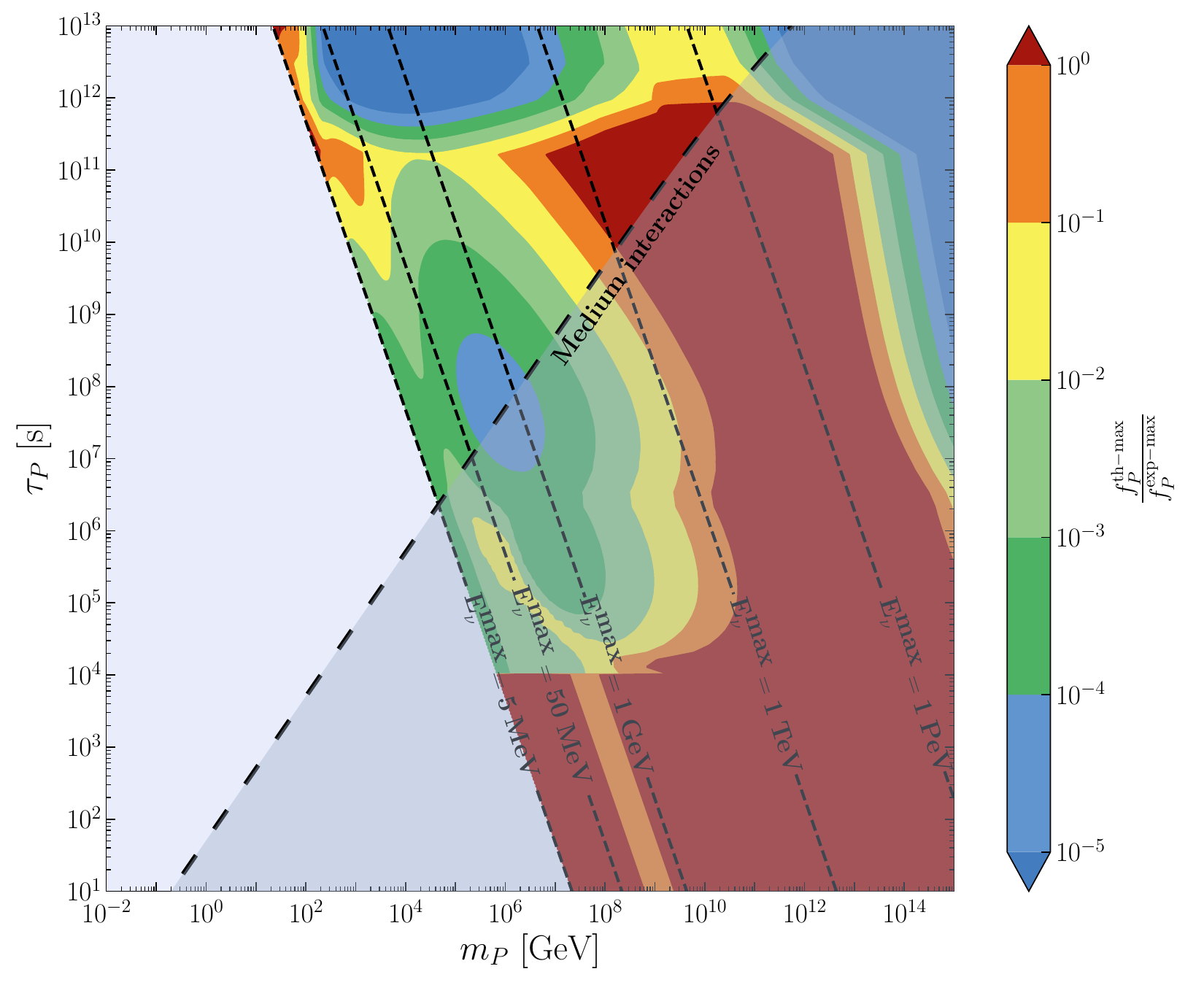}
    \includegraphics[width=0.5425\linewidth]{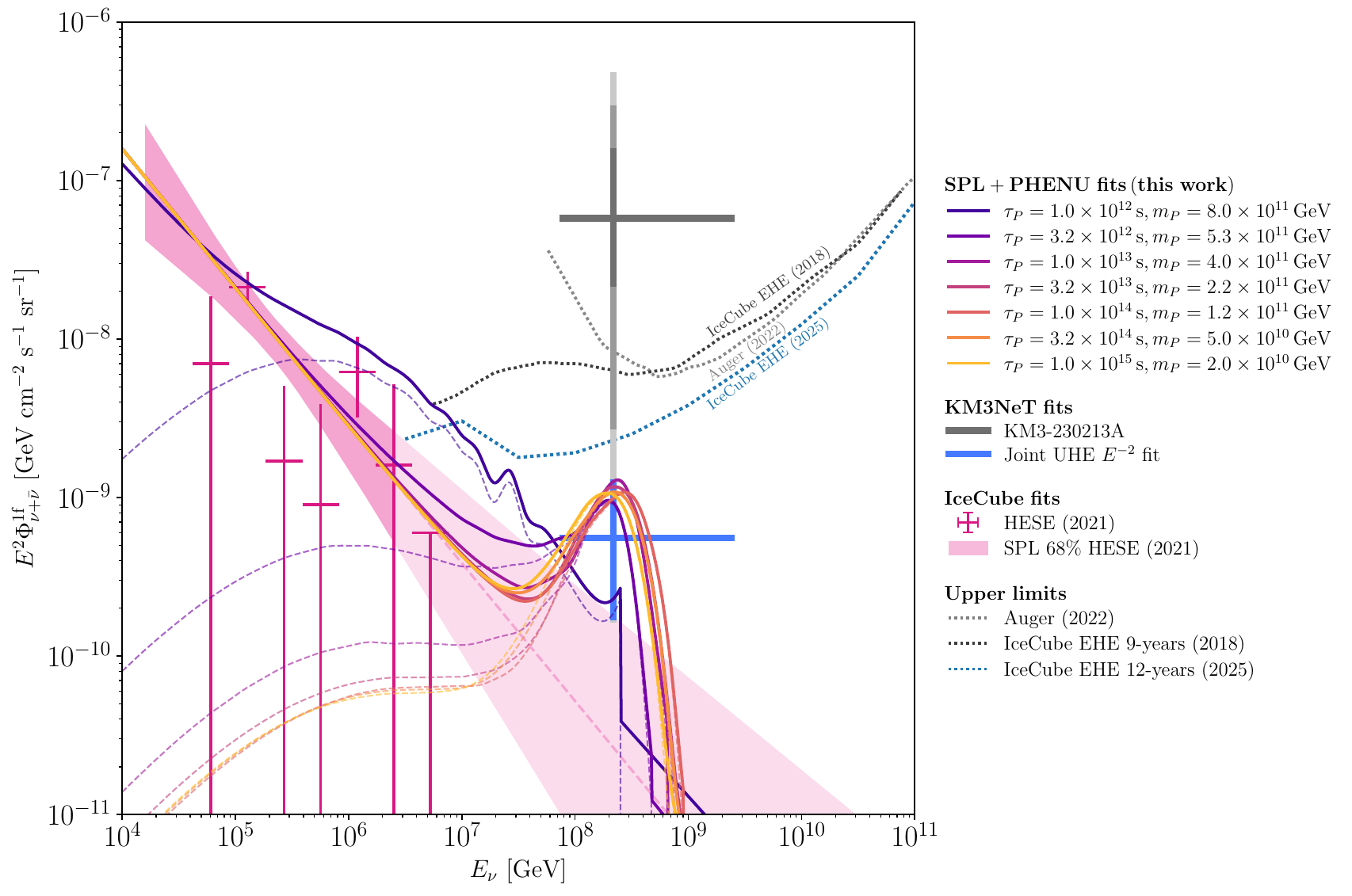}
    \caption{\emph{Left}: Ratio of the maximal $f_P$ allowed by the CMB, BBN and
$\Delta N_{\rm eff}$ constraints to the one allowed by experimental constraints, in
the $(m_P, \tau_P)$ plane. The light grey region corresponds to parameters giving
$E_\nu^{\rm max} < 5$\,MeV, difficult to probe experimentally. The dashed black line
delimits the region where the produced Phenu undergoes on average less than one elastic
scattering with a C$\nu$B neutrino on its way to the detector, so that the sharp
spectral feature is safely undistorted (unshaded region). \emph{Right}: Best-fit Phenu
spectra from the joint fit to KM3NeT, IceCube HESE, IceCube EHE and Pierre Auger data,
for seven values of $\tau_P$ ranging from $10^{12}$ to $10^{15}$\,s, with $m_P$ fixed
in each case so that the spectral peak matches the KM3-230213A event energy
($\sim 200$\,PeV). More details can be found in~\protect\cite{GY:2026}.}
    \label{fig:parameter_space}
\end{figure}

For the two-body decay case, the source particle is characterised by its mass $m_P$,
lifetime $\tau_P$, and abundance $f_P \equiv \Omega_P^0/\Omega_{\rm DM}^0$ with $\Omega^0_P$ and $\Omega^0_{\rm DM}$ the fraction of the total energy density today of, respectively, the heavy relic $P$ if it did not decay and that of dark matter.
A first crucial question is whether in-flight scatterings of Phenus on C$\nu$B
neutrinos are negligible, since elastic and inelastic interactions can distort or
entirely erase the sharp spectral feature. We determine the region of $(m_P, \tau_P)$
parameter space for which the average number of such scatterings is less than one,
so that the spectral feature arrives undistorted.
Within this region (above the shaded grey area in the left panel of figure~\ref{fig:parameter_space}), we then apply the theoretical constraints: the $\Delta N_{\rm eff}$
bound from the CMB, which limits the energy injected into radiation before
recombination; and CMB anisotropy and BBN photo- and hadrodisintegration constraints,
the latter taken from Ref.~\cite{Bianco:2025} and the former recast from
Ref.~\cite{Acharya:2019}, which are stringent for $\tau_P \gtrsim 10^4$\,s and $m_P$
of order 100 GeV or above.

Combining these with experimental upper bounds from atmospheric and astrophysical
neutrino flux measurements by Super-Kamiokande and IceCube, we identify an
observable and distortion-free region spanning source particle masses from a few tens
of GeV up to $\sim 10^{11}$\,GeV, with lifetimes between $\sim 10^9$\,s and
recombination ($\sim 10^{13}$\,s). The corresponding Phenu energies today range from
$\sim 10$\,GeV to a few PeV, fully accessible to IceCube, KM3NeT and
Hyper-Kamiokande. This is shown in the left panel of figure~\ref{fig:parameter_space},
where the colour scale gives the ratio of the maximal abundance allowed by theoretical
constraints to that allowed by experimental ones, indicating by how much a Phenu flux
could saturate the observed neutrino flux. The orange and red regions correspond
to the most promising part of the parameter space, where a Phenu signal could
potentially be within reach of current neutrino telescopes.

\section{Monte Carlo code for spectral distortion}

Within the grey shaded area of the parameter space shown in the left panel of
figure~\ref{fig:parameter_space}, interactions with the C$\nu$B become frequent and a
proper computation of the distorted spectrum is necessary. At these energies, final
state radiation (FSR) at production is also non-negligible. Both effects are relevant
for the $\sim 200$\,PeV KM3-230213A event if interpreted as a Phenu~\cite{KM3NeT:2025}.

We have developed a dedicated Monte Carlo code for this purpose. The simulation tracks
individual neutrinos produced by the decay of a heavy relic along adaptive redshift
steps. At each step, it accounts for relic decay with FSR via
HDMSpectra~\cite{HDMSpectra}, elastic and inelastic scatterings with the C$\nu$B,
redshift energy loss, and flavour oscillations.

Using this code, we perform a joint fit of a signal model consisting of an
astrophysical power-law background and a Phenu component to KM3NeT, IceCube HESE,
IceCube EHE and Pierre Auger data~\cite{GY:2026}. The fit is run independently for
seven benchmark values of $\tau_P$ ranging from $10^{12}$ to $10^{15}$\,s, with $m_P$
fixed in each case so that the spectral peak matches the KM3-230213A event energy.
Six of the seven spectra reduce the tension between the UHE datasets, bringing it from
$3.1\sigma$ for a pure power law down to $\approx 2.85\sigma$, close to the irreducible
$2.7\sigma$ tension arising from the non-observation of events at the same energy by
IceCube and Auger. The best-fit spectra are shown in the right panel of
figure~\ref{fig:parameter_space}, overlaid on the data from all four experiments. 
Remarkably, the best-fit Phenu abundance is found to be close to the upper bound
imposed by CMB constraints, suggesting that future CMB experiments could also be
sensitive to this scenario. The full results are reported in Ref.~\cite{GY:2026}.

\section{Conclusion}

Phenus offer a novel probe of the early Universe through high energy neutrino
observations. A general study~\cite{GY:2025} shows that a wide region of source
particle parameter space is viable and that sharp spectral features can survive
propagation unaltered. Extending this to regimes where C$\nu$B interactions and FSR
are non-negligible, a joint fit to KM3NeT, IceCube and Pierre Auger data shows that the KM3-230213A event can be accommodated within the Phenu scenario,
reducing the UHE dataset tension. Remarkably, the best-fit abundance lies close to the CMB upper
bound, suggesting future CMB experiments could also be sensitive to Phenus.

\section*{Acknowledgments}

This work is based on a collaboration with Thomas Hambye, refs.~\cite{GY:2025,GY:2026}. The work of N.G.Y.\ is supported by
the Communaut\'e fran\c{c}aise de Belgique through a FRIA doctoral grant.
Computational resources were provided by the CECI consortium, funded by the F.R.S.-FNRS
under Grant No.\ 2.5020.11 and by the Walloon Region. N.G.Y is a member of BLU-ULB (Brussels Laboratory of the Universe, blu.ulb.be). Report number ULB-TH/26-19.

\section*{References}
\bibliography{moriond}

\end{document}